\documentclass[a4paper,12pt]{article}
\usepackage{graphicx}
\global\arraycolsep=2pt %reduces the separation in eqnarrays
\input{epsf}
\begin{document}
%-----------------------------
\makeatletter
\def\fmslash{\@ifnextchar[{\fmsl@sh}{\fmsl@sh[0mu]}}
\def\fmsl@sh[#1]#2{%
  \mathchoice
    {\@fmsl@sh\displaystyle{#1}{#2}}%
    {\@fmsl@sh\textstyle{#1}{#2}}%
    {\@fmsl@sh\scriptstyle{#1}{#2}}%
    {\@fmsl@sh\scriptscriptstyle{#1}{#2}}}
\def\@fmsl@sh#1#2#3{\m@th\ooalign{$\hfil#1\mkern#2/\hfil$\crcr$#1#3$}}
\makeatother
%--------------------------------
%---------------- CERN Titlepage <---------------------------
\thispagestyle{empty}
\begin{titlepage}

\begin{flushright}
hep-ph/0008252 \\
LMU 11/00 \\
\today
\end{flushright}

\vspace{0.3cm}
\boldmath
\begin{center}
  \Large \bf The Higgs Boson Might Not Couple \\
  {\Large {\bf To $B$ Quarks.}}
\end{center}
\unboldmath
\vspace{0.8cm}
\begin{center}
  {\large Xavier Calmet}\\
  
\end{center}
\begin{center}
  and
  \end{center}
\begin{center}
{\large Harald Fritzsch}\\
 \end{center}
 \vspace{.3cm}
\begin{center}
{\sl Ludwig-Maximilians-University Munich, Sektion Physik}\\
{\sl Theresienstra{\ss}e 37, D-80333 Munich, Germany}\\
\end{center}

\vspace{\fill}

\begin{abstract}
\noindent 
We discuss an alternative version of the electroweak standard model,
in which only the heavy $t$ quark, not the light fermions, couples to
the Higgs boson with a strength given by the standard model. The Higgs
particle decays dominantly into two gluons jets. The branching ratio
for the $2 \, \gamma$ decay is about $3.5 \, \%$. The Higgs particle
would be a narrow object (width about $60$ KeV), and its mass might be
consistent with the value given by typical estimates of radiative
effects measured by the LEP experiments.
\end{abstract}
to appear in Physics Letters B.
\end{titlepage}
% ----------------------------------------------------------
As far as the mass generation within the framework of the standard
electroweak model is concerned, one must differentiate between the
mass generation for the electroweak bosons $W$, $Z$, the mass
generation for the heavy $t$ quark, and the generation of mass for the
leptons and the five remaining, relatively light quarks. While there
exists no freedom in the choice of the interaction strengths of the
weak bosons with the scalar field, which is dictated by the gauge
invariance \cite{LlewellynSmith:1973ey}, there is such a freedom with
respect to the fermions. The masses of the fermions are given by the
various Yukawa coupling constants, which parametrize the interactions
of the leptons and quarks with the scalar field. The Yukawa coupling
constant of the $t$ quark field is of the same order as the gauge
coupling constant, while the other fermions couple much weakly ($0.018$
for the $b$ quark, $0.005$ for the $c$ quark, etc.). The origin of the
light fermion masses is still mysterious, and alternative views or
slight variations of the standard electroweak theory might indeed give
a different view.  Taking into account the observed flavor mixing
phenomenon, one could speculate, for example, that the masses of the
light quarks and of the leptons are due to the mixing. In the absence
of the mixing the mass matrix of the quarks in the $u$-sector would
simply be proportional to a diagonal matrix with the entries $(0, 0,
1)$, and there would only be a coupling of the scalar field to the $t$
quark. Once the flavor mixing is switched on, the mass eigenstates for
the light quarks are not necessarily coupled to the scalar field, with
a strength given by the mass eigenvalues. In particular these
couplings could remain zero.

It is well-known that the renormalizability of the theory requires a
coupling of the fermion to the scalar field
\cite{LlewellynSmith:1973ey}. Otherwise the unitarity in the $s$
channel is violated at high energies for the reaction $f_L \, \bar f_L
\rightarrow W^+ W^-$.

However, for all fermions except the $t$ quark these problems appear
only at extremely high energies. Modifications of the electroweak
theory, which involve an energy scale not orders of magnitude above
the typical electroweak scale of about $0.3$ TeV, e.g. theories which
do not rely on the Higgs mechanism, can take care of this problem.

Recently we have discussed such a modification \cite{CF1}, or rather
an alternative description of the standard model, based on the
complementarity between confinement and Higgs phase
\cite{Fradkin:1979dv}. We suppose that the electroweak interactions
are described by the confinement phase, and not by the Higgs phase, as
usually assumed. This provides an alternative view of the electroweak
bosons, which are not the basic gauge bosons of the underlying gauge
theory, but ``bound states'' of an underlying scalar field, which in
the Higgs phase plays the role of the Higgs doublet.  Both the charged
$W$ bosons and the neutral $Z$ boson are $J = 1$ bound systems of the
type $hh, (hh)^\dagger$ or $\bar h h$ respectively.  There is a
corresponding $J = 0$, $\bar h h$ system, which is to be identified
with Higgs boson of the standard electroweak model.

In a simple non-relativistic picture of these bound states the $s$
wave state would in general have a mass less than the mass of the $p$
wave state, e.g. the Higgs bosons would have a mass below $81$ GeV.
This need not be the case here, due to relativistic effects, and due
to the complicated interplay between the confining gauge force and the
scalar self-interaction, but nevertheless the mass of the Higgs boson
is not expected to be very large compared to the $W$ boson. The mass
splitting between the Higgs boson and the $W$ boson could be
calculated using the lattice approximation, but since this has not yet
been done, we are not able to make predictions for the Higgs mass.

We shall consider a deviation from our original model which would have
the same couplings as in the standard model.  It is conceivable that
in the confinement phase of the electroweak theory the coupling
strength of the fermions to the scalar boson are not proportional to
the light fermion masses, since these couplings depend strongly on the
dynamics of the model. In the simplest case only the fermion whose
mass is of the same order as the weak interaction energy scale, i.e.,
the $t$ quark, has such a coupling. Thus we proceed to calculate the
properties of the scalar boson, which couples only to the $t$ quark.
As far as the interaction of such a boson with the $W$ and $Z$ bosons
is concerned, there is no change in comparison to the standard
electroweak model.  However there is a substantial change of the decay
properties. Decay modes which were regarded as being strongly
suppressed become dominant.

\begin{figure}[ht]
  \begin{minipage}[t]{0.49\linewidth}\centering
    \includegraphics[width=\linewidth]{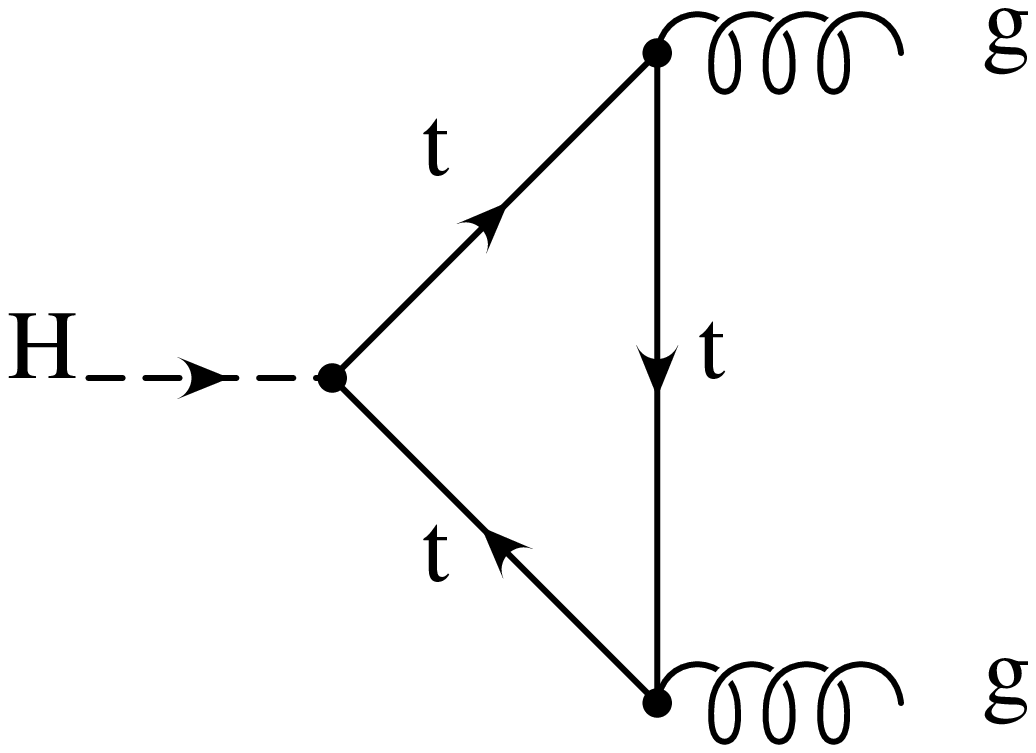}
    \begin{minipage}{0.9\linewidth}
      \caption{Top triangle.
        \label{graph1}}
    \end{minipage}
  \end{minipage}
  \begin{minipage}[t]{0.49\linewidth}\centering
    \includegraphics[width=\linewidth]{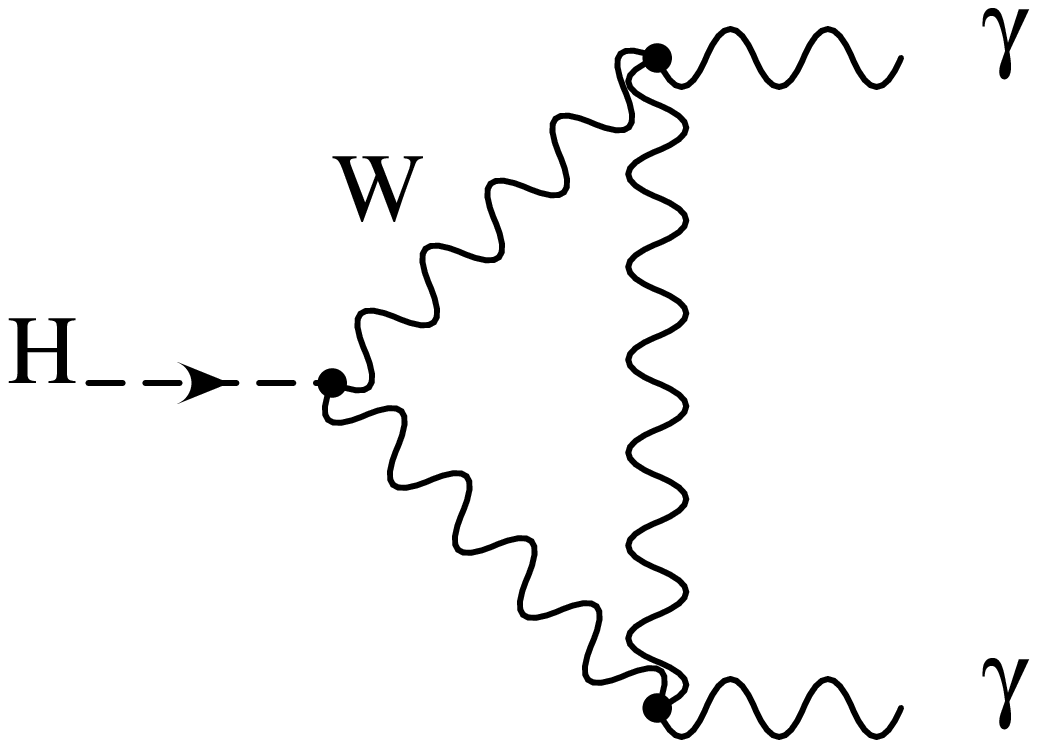}
    \begin{minipage}{0.9\linewidth}
      \caption{$W$ triangle.
        \label{graph2}}
    \end{minipage}
  \end{minipage}
  \begin{minipage}[t]{0.49\linewidth}\centering
    \includegraphics[width=\linewidth]{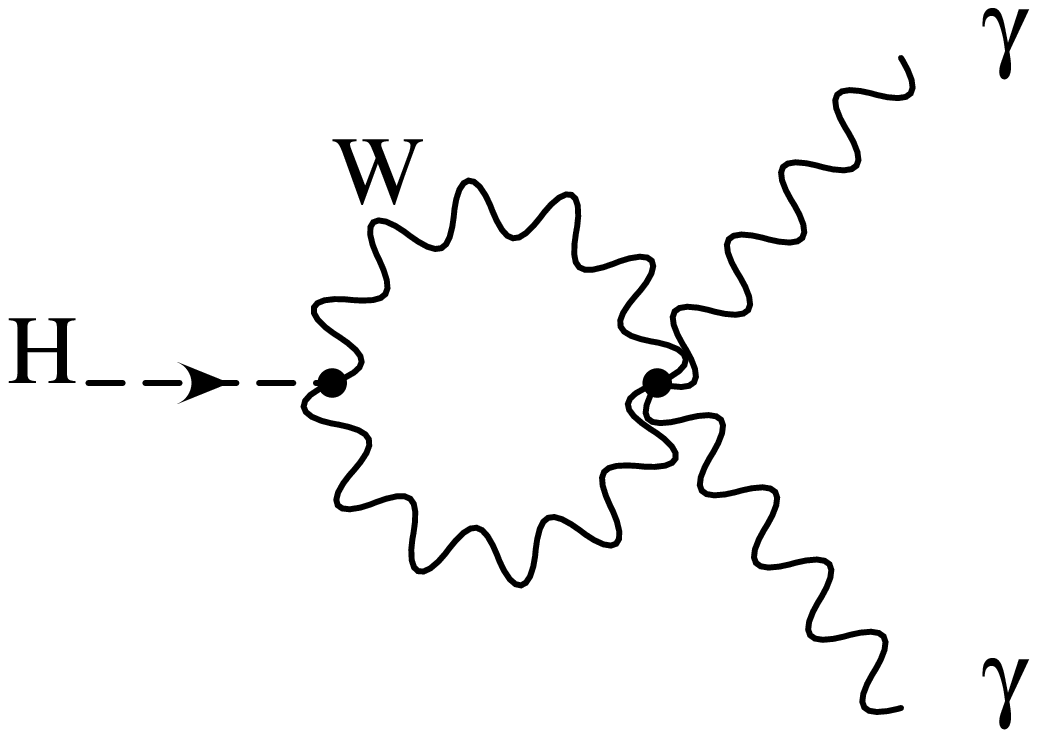}
    \begin{minipage}{0.9\linewidth}
      \caption{$W$ bubble.
        \label{graph3}}
    \end{minipage}
  \end{minipage}
  \begin{minipage}[t]{0.49\linewidth}\centering
    \includegraphics[width=\linewidth]{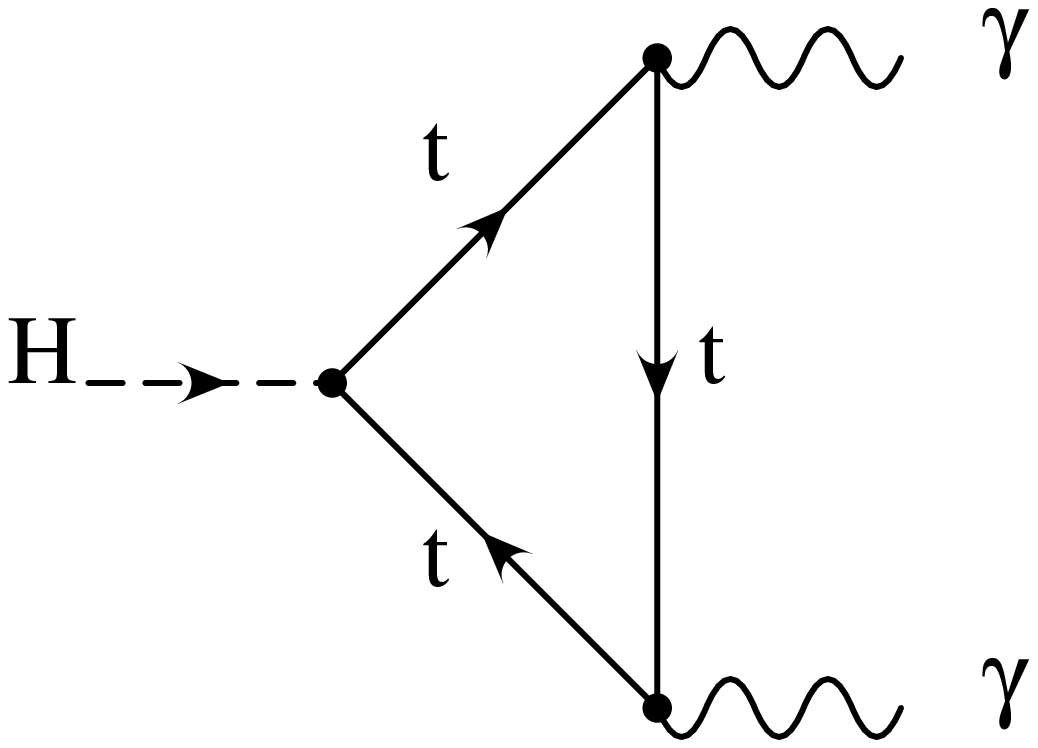}
    \begin{minipage}{0.9\linewidth}
      \caption{Top triangle.
        \label{graph4}}
    \end{minipage}
  \end{minipage}
  \begin{minipage}[t]{0.49\linewidth}\centering
    \includegraphics[width=\linewidth]{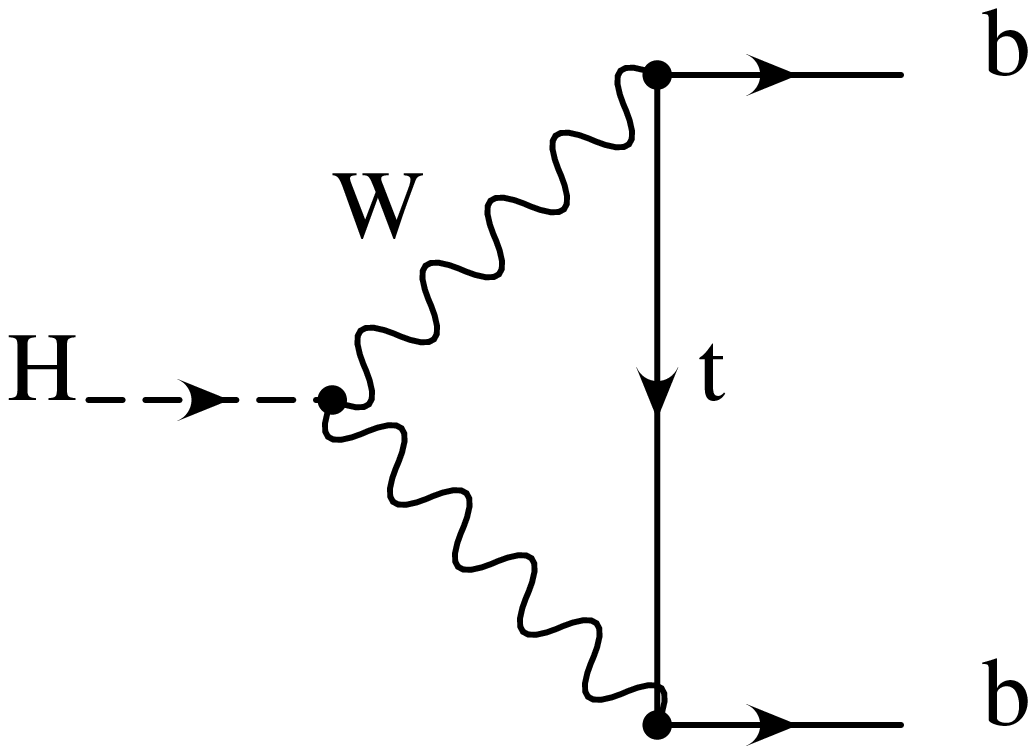}
    \begin{minipage}{0.9\linewidth}
      \caption{1st effective b quark decay.
        \label{graph5}}
    \end{minipage}
  \end{minipage}
  \begin{minipage}[t]{0.49\linewidth}\centering
    \includegraphics[width=\linewidth]{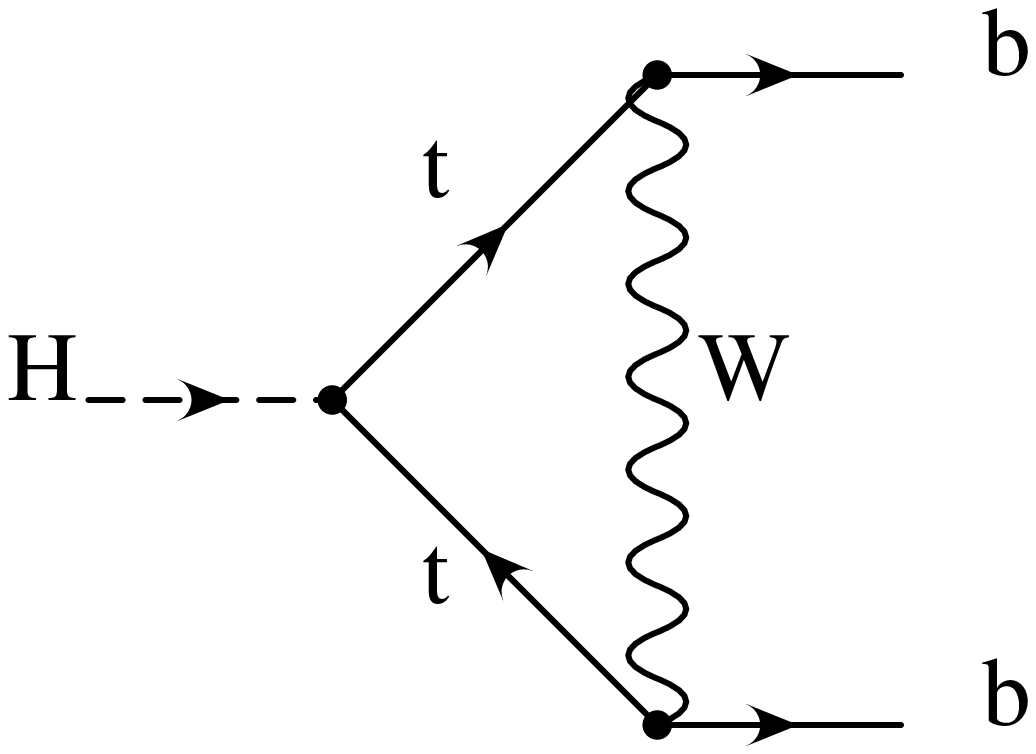}
    \begin{minipage}{0.9\linewidth}
      \caption{2nd effective b quark decay.
        \label{graph6}}
    \end{minipage}
  \end{minipage}
\end{figure}

We thus consider the following decay channels for the Higgs boson:
$H \to g g$ (see graph \ref{graph1}) via a top quark triangle and
$H \to \gamma \gamma$ (see graphs \ref{graph2}, \ref{graph3} and
\ref{graph4}) via a triangle involving top quarks and charged
electroweak bosons or a bubble diagram involving a neutral electroweak
boson.  For a two photon Higgs decay, ignoring radiative corrections,
one finds \cite{Ellis:1976ap}
\begin{eqnarray}
\Gamma(H \to \gamma \gamma) &=& \frac{ \alpha^2 g^2}{1024 \pi^3} 
\frac{M_H^3}{M_W^2} \left | \sum_i e_i^2 N_{c\, i} F_i \right |^2 
= \frac{ \alpha^2 g^2}{1024 \pi^3} 
\frac{M_H^3}{M_W^2} \left | \frac{4}{3} F_{1/2} + F_{W} \right |^2 
\end{eqnarray}
where the functions $F_{1/2}$ and $F_{W}$ 
are given by
\begin{eqnarray} \label{eq1}
F_{1/2}&=&-2 \tau [1+(1-\tau) f(\tau)]
\end{eqnarray}
and
\begin{eqnarray}
F_{W}&=& 2+3 \tau + 3 \tau (2-\tau) f(\tau)
\end{eqnarray}
where $\tau=4m_i^2/M_H^2$.  The first function corresponds to the
contribution of the top quark and the second to the contribution of
the charged $W$ bosons.  As we assume that the Higgs boson is light,
i.e., lighter than twice the mass of the $W$ bosons, the function
$f(\tau)$ reads
 \begin{eqnarray}
 f(\tau)&=& \left ( \arcsin{ \left( \sqrt{\frac{1}{\tau}}\right )} \right )^2.
 \end{eqnarray}  
 For the decay into two gluons one finds \cite{Ellis:1976ap}
 \begin{eqnarray} 
\Gamma(H \to g g) &=& \frac{ \alpha_s^2 g^2}{512 \pi^3}
\frac{M_H^3}{M_W^2} \left |  F_{1/2} \right |^2, 
\end{eqnarray}
also neglecting the radiative corrections. The function $F_{1/2}$ was
given in equation (\ref{eq1}).

Another possibility for the Higgs boson to decay are the electroweak
boson channels $H \to W W$ and $H \to Z Z$. The Higgs boson couples to
the electroweak bosons with the same strength as in the standard model.
The decay via two virtual electroweak bosons represents a
non-negligible contribution to the Higgs decay. For $m_W<m_H$ or
$m_Z<m_H$ one of the electroweak bosons is on-shell.  These decay
rates were evaluated using the program HDECAY \cite{Djouadi:1998yw}
and cross-checked using CompHEP \cite{Pukhov:1999gg}. The numerical
results are the sum of the decay over two electroweak bosons, for a
light Higgs both electroweak bosons are virtual, when allowed by the
kinematics, the contributions of on-shell electroweak bosons are also
taken into account.

\begin{table}
\centering
  \begin{tabular}{|l|l|l|l|l|l|}
   \hline
   channel & $\!m_H\!=\!60$ 
   & $\!m_H\!=\!70$ 
   & $\!m_H\!=\!80$ 
   & $\!m_H\!=\!90$ 
   & $\!m_H\!=\!100$
   \\
    \hline
   $\Gamma(H \! \to \! g  g)_{\alpha_s=0.119} \! \! 
   $ &$2.3\times 10^{-5}  $    &
  $3.7\times 10^{-5}$ & $5.5\times 10^{-5}$ 
  & $7.9\times 10^{-5}$  
  &$1.1\times 10^{-4}$    \\
 \hline
  $\Gamma(H \! \to \! g  g)_{\alpha_s=0.15} \! \! 
  $ &$3.6\times 10^{-5}$ & 
  $5.8\times 10^{-5}$ & $8.7\times 10^{-5}$ 
  & $1.3\times 10^{-4}$  
  &$1.7\times 10^{-4}$    \\
 \hline
 $\Gamma(H \!\to \! \gamma \, \gamma )$
 &$8.0\times 10^{-7}$ &
 $1.3\times 10^{-6}$&
 $2.0\times 10^{-6}$
 &$3.0\times 10^{-6}$ &
 $4.4\times 10^{-6}$
 \\
 \hline
 $\Gamma(H\! \to \!W \, W)\! \! 
 $ &$1.09\times 10^{-7}$ &$3.80\times 10^{-7}$ &
 $1.22\times 10^{-6} $
 &$4.49\times 10^{-6} $ &$2.66\times 10^{-5} $ \\
 \hline
  $\Gamma(H \! \to \! Z \, Z)\! \! 
  $ &$3.33\times 10^{-8}$ &
  $1.10\times 10^{-7} $ &
  $3.27\times 10^{-7} $
  &
  $9.12\times 10^{-7} $ &
 $2.72\times 10^{-6} $  \\
 \hline
\end{tabular}
 \caption{Higgs boson decay rates in GeV for different Higgs masses in GeV.
         \label{tab:table1}}
 \end{table}
 
 The results of these calculations are given in table
 \ref{tab:table1}.  The corresponding branching ratios are given in
 table \ref{tab:table2}. We see that such a Higgs boson would decay in
 a fundamentally different way than the Higgs boson of the standard
 model. The results for the $H \to g g$ decay are strongly dependent
 of the value chosen for $\alpha_s$. Thus this decay channel has a
 considerable uncertainty.  We have done the calculations for two
 different values of the strong coupling constant $\alpha_s=0.119$ and
 $\alpha_s=0.15$. The fine-structure constant was taken to be
 $\alpha=1/128.9$.
 
 Even if the light fermions in particular the $b$ quark, do not couple
 directly to the Higgs boson, some $b$ quarks could be produced via
 the diagrams \ref{graph5} and \ref{graph6}. Their contributions is
 not easy to estimate but the electroweak corrections for a light
 Higgs are known to be very small \cite{Kniehl:1992ze}, typically $0.3
 \%$ of the tree level value.  Nevertheless they could still be of the
 same order of magnitude as the $\gamma \gamma$ contribution.  Above
 $90$ GeV the decay channel $H \to Z \gamma$ opens. For masses larger
 than $110$ GeV the Higgs boson mainly decays into two electroweak
 bosons.

 The present searches for the Higgs boson at LEP are mainly based on
 the assumption that the leading decay made in the mass region of
 about 100 GeV or less is the decay $H \rightarrow \bar b b$. The
 present experimental limit $m_H >113.3$ GeV \cite{Murray} is obtained
 on the basis of this assumption. In our model the decay is dominated
 by the decay $H \rightarrow g g$, i.e., the decay products do not show
 a specific flavor dependence. The lower limit on the mass of such a
 boson is much weaker and of the order of $70$ GeV \cite{Schaile}.
 
 The best way to detect the Higgs boson at LEP seems to us to search
 for the decay $H \rightarrow \gamma \gamma $. Since the invariant
 mass of the $2 \gamma $ system would be identical to the mass of the
 boson, the background coming from radiation effects could be
 substantially reduced. In our case this decay channel, having a small
 branching ratio, is not seriously constrained by fermiophobic Higgs
 studies \cite{Murray}.

 Typical fits of the Higgs boson mass indicate that the most likely
 mass of the boson is about $77^{+69}_{-39}$ GeV \cite{PDG}. It might
 well be, that the mass of the Higgs boson is in the region $70$ to
 $110$ GeV, provided the decay proceeds via the mechanism discussed
 above. We note that in contrast to the standard expectation the Higgs
 particle is a relatively narrow object with a width of about $58.5$
 KeV.
\begin{table}[ht]
\centering
  \begin{tabular}{|l|l|l|l|l|l|}
   \hline
   channel & $\!m_H\!=\!60$ 
   & $\!m_H\!=\!70$ 
   & $\!m_H\!=\!80$ 
   & $\!m_H\!=\!90$ 
   & $\!m_H\!=\!100$
   \\
    \hline
   ${\mbox Br}(H \! \to \! g  g)$
   & $96.06 \, \%$
   & $95.39 \, \%$
   & $93.94 \, \%$
   & $90.39 \, \%$ 
   & $76.54 \, \% $\\
 \hline
 ${\mbox Br}(H \!\to \! \gamma \, \gamma )$ &$3.34\, \%  $ &
 $3.35 \, \%$&
 $3.42 \, \%$
 &$3.43 \, \%$
 & $3.06\, \% $
 \\
 \hline
 ${\mbox Br}(H\! \to \!W \, W)$ &$0.46 \, \% $
 &$0.98 \, \% $
 & $2.08 \, \% $
 &$5.14 \, \% $
& $18.51\, \% $
 \\
 \hline
  ${\mbox Br}(H \! \to \! Z \, Z)$ &$0.14 \, \% $ &
  $0.28 \, \% $ &
  $0.56 \, \% $
  &
  $1.04 \, \% $
& $1.89 \, \% $
  \\
 \hline
\end{tabular}
 \caption{Branching ratios for different Higgs masses in GeV and
   for $\alpha_s=0.119$. \label{tab:table2}}
 \end{table}

\section*{Acknowledgements}
We should like to thank G. Duckeck, B. Kniehl, A. Leike, A. Martin, M.
Stanitzki and P. Zerwas for useful discussion.

\end{document}